\documentclass[aps,twocolumn,nofootinbib,superscriptaddress]{revtex4}

\usepackage{amsfonts,amsthm}

\renewcommand{\>}{\rangle}

\newcommand{\be}{\begin{equation}}
\newcommand{\ee}{\end{equation}}
\newcommand{\bea}{\begin{eqnarray}}
\newcommand{\eea}{\end{eqnarray}}

\newcommand{\tp}{\otimes}
\newcommand{\nn}{\nonumber\\}

\newtheorem{theorem}{Theorem}
\newtheorem{cor}[theorem]{Corollary}
\newtheorem{lemma}[theorem]{Lemma}

\begin{document}

\title{Lower bounds on the complexity of simulating quantum gates}

\author{Andrew M. Childs}
\email[]{amchilds@mit.edu}
\affiliation{Center for Theoretical Physics,
             Massachusetts Institute of Technology,
             Cambridge, MA 02139, USA}

\author{Henry L. Haselgrove}
\email[]{hlh@physics.uq.edu.au}
\affiliation{School of Physical Sciences,
             The University of Queensland,
             Brisbane QLD 4072, Australia}
\affiliation{Information Sciences Laboratory,
             Defence Science and Technology Organisation,
             Edinburgh 5111, Australia}

\author{Michael A. Nielsen}
\email[]{nielsen@physics.uq.edu.au}
\homepage[]{www.qinfo.org/people/nielsen}
\affiliation{School of Physical Sciences,
             The University of Queensland,
             Brisbane QLD 4072, Australia}
\affiliation{School of Information Technology and Electrical Engineering,
             The University of Queensland,
             Brisbane QLD 4072, Australia}
\affiliation{Institute for Quantum Information,
             California Institute of Technology,
             Pasadena CA 91125, USA}

\date[]{25 July 2003}


\begin{abstract}
  We give a simple proof of a formula for the minimal time required to
  simulate a two-qubit unitary operation using a fixed two-qubit
  Hamiltonian together with fast local unitaries.  We also note that a
  related lower bound holds for arbitrary $n$-qubit gates.
\end{abstract}

\maketitle

\section{Introduction}

%
%
Understanding quantum dynamics is at the heart of quantum physics.
Recent ideas from quantum computation have stimulated interest in
studying the physical resources needed to implement quantum
operations.  In addition to a qualitative understanding of what
resources are necessary, we would like to quantify the resource
requirements for universal quantum computation and other information
processing tasks.  Ultimately, we would like to understand the minimal
resources that are necessary and sufficient to implement particular
quantum dynamics.

As a first step towards answering these questions, it has been shown
that there is a sense in which all entangling dynamics are
qualitatively equivalent.  In particular, it has been shown that any
$n$-qudit two-body Hamiltonian capable of creating entanglement
between any pair of qudits is, in principle, universal for quantum
computation, when assisted by arbitrary single-qudit unitaries
\cite{Jones99a,Leung00a,Dodd02a,Wocjan02a,Dur01a,Bennett01a,Nielsen02f,Vidal01b}.
Thus, any particular entangling two-qudit Hamiltonian can be used to
simulate any other, provided local unitaries are available.  This
suggests that such dynamics are a fungible physical resource.

Having established the qualitative equivalence of all entangling
dynamics, we would like to quantify their information processing
power.  In particular, it is interesting to consider the minimal time
required to implement a unitary operation, $U$, on a two-qubit system,
using a fixed Hamiltonian, $H$, and the ability to intersperse fast
local unitary operations on the two qubits.  This problem was studied
by Khaneja, Brockett and Glaser~\cite{Khaneja01a}, who found a
solution using the theory of Lie groups.  Their results, although
giving a solution in principle, are neither explicit about the form of
the minimal time, nor do they explain how to construct all elements of
the time-optimal simulation.  Further work by Vidal, Hammerer, and
Cirac \cite{Vidal02a}, from a different point of view, resulted in an
explicit formula for the minimal time, and gave a constructive
procedure for minimizing that time (see also~\cite{Hammerer02a}, where
an alternate proof is given by the same authors).

%
%
The purpose of the present paper is to give a simplified proof that
the formula of Vidal, Hammerer and Cirac is, in fact, a lower bound on
the simulation time.  Note that the difficult part
of~\cite{Vidal02a,Hammerer02a} was proving the lower bound; finding a
protocol to meet the lower bound was comparatively easy.

%
%
The main advantages of our proof are its simplicity and conceptual
clarity, as compared to the ingenious, but rather complex, arguments
in~\cite{Khaneja01a,Vidal02a,Hammerer02a}.  This simplicity is
achieved by making use of a powerful result from linear algebra,
Thompson's theorem.  We expect that Thompson's theorem might be useful
for many other problems in quantum information theory.  A second
advantage of using Thompson's theorem is that it does not rely on
special properties of two-qubit unitary operators.  Therefore,
essentially the same arguments give a lower bound on the time required
to implement an $n$-qubit unitary operation using a fixed $n$-qubit
interaction Hamiltonian, and fast local unitary operations.

%
%
Our approach to the proof of the lower bound has its roots in the
framework of \emph{dynamic strength measures} for quantum
operations~\cite{Nielsen03a}.  The dynamic strength framework is an
attempt to develop a quantitative theory of the power of dynamical
operations for information processing.  The idea is to associate with
a quantum dynamical operation, such as a unitary operation $U$, a
quantitative measure of its ``strength.'' In~\cite{Nielsen03a} it was
shown that such strength measures can be used to analyze the minimal
time required for the implementation of a quantum operation.  The
present paper takes a similar approach, but instead of using a single
real number to quantify dynamic strength, we use a vector-valued
measure.  This can also be compared to the analysis of optimal
simulation of Hamiltonian dynamics using a set of several strength
measures \cite{Childs03b}.

%
%
Our paper is structured as follows.  Section~\ref{sec:background}
reviews some background material on majorization, Thompson's theorem,
and the structure of the two-qubit unitary matrices.  The main result,
the lower bound on optimal simulation, is proved in
Section~\ref{sec:two-qubit}.  We conclude in Section~\ref{sec:conc} by
presenting our generalization of the lower bound to $n$ qubits and
suggesting some directions for future work.  In addition, an appendix
gives a procedure for calculating a canonical decomposition of
two-qubit unitary gates.

\section{Background}
\label{sec:background}

This section reviews the relevant background needed for our proof.
Section~\ref{subsec:majorization} reviews the basic notions of
majorization, introduces Thompson's theorem, and explains how to use
Thompson's theorem and majorization to relate properties of a product
of unitary operators to properties of the individual unitaries.
Section~\ref{subsec:canonical} introduces the \emph{canonical
decomposition}, a useful representation theorem for two-qubit unitary
operators, and Section~\ref{subsec:canonicalham} presents an analogous
decomposition for Hamiltonians.

\subsection{Majorization and Thompson's theorem}
\label{subsec:majorization}

%
%
Our analysis uses the theory of majorization together with Thompson's
theorem.  More detailed introductions to majorization may be found in
\cite{Nielsen01b}, Chapter~2 and~3 of~\cite{Bhatia97a}, and
in~\cite{Marshall79a,Alberti82a}.

%
%
Suppose $x = (x_1,\ldots,x_D)$ and $y = (y_1,\ldots,y_D)$ are two
$D$-dimensional real vectors.  The relation \emph{$x$ is majorized by
$y$}, written $x \prec y$, is intended to capture the intuitive notion
that $x$ is \emph{less ordered} (i.e., more disordered) than $y$.  To
make the formal definition we introduce the notation $\downarrow$ to
denote the components of a vector rearranged into non-increasing
order, so $x^{\downarrow} =
(x^{\downarrow}_1,\ldots,x^{\downarrow}_D)$, where $x^{\downarrow}_1
\geq x^{\downarrow}_2 \geq \ldots \geq x^{\downarrow}_D$.  Then $x$ is
majorized by $y$, that is, $x \prec y$, if
\be
  \sum_{j=1}^k x^{\downarrow}_j \leq \sum_{j=1}^k y^{\downarrow}_j
\ee
for $k = 1,\ldots,D-1$, and the inequality holds with equality when $k
= D$.

%
%
To connect majorization to Hamiltonian simulation, we use a result of
Thompson relating a product of two unitary operators to the individual
unitary operators.  Recall that an arbitrary pair of unitary operators can
be written in the form $e^{iH}$ and $e^{iK}$, for some Hermitian $H$ and
$K$.  Thompson's theorem provides a representation for the product
$e^{iH}e^{iK}$ in terms of $H$ and $K$:
\begin{theorem}[Thompson~\cite{Thompson86a}]\label{thm:thompson} Let $H,K$
be Hermitian matrices.  Then there exist unitary matrices $U,V$ such that
\be
  e^{i H} e^{i K} = e^{i(U H U^\dag + V K V^\dag)}.
\ee
\end{theorem}

%
%
The proof of Thompson's theorem in~\cite{Thompson86a} depends on a
result conjectured earlier by Horn~\cite{Horn62a}.  A proof of this
conjecture had been announced and outlined by
Lidskii~\cite{Lidskii82a} at the time of Thompson's paper.  However,
remarks in~\cite{Thompson86a} suggest that~\cite{Lidskii82a} did not
contain enough detail to be considered a fully rigorous proof.
Fortunately, a proof of Horn's conjecture has recently been fully
completed and published.  See, for
example,~\cite{Fulton00a,Knutson00a} for reviews and references.

%
%
Thompson's theorem may be related to majorization using the following
theorem of Ky Fan:
\begin{theorem}[Ky Fan \cite{Fan49a,Bhatia97a}]\label{thm:fan}
  Let $H,K$ be Hermitian matrices.  Then $\lambda(H+K) \prec
  \lambda(H)+\lambda(K)$, where $\lambda(A)$ denotes the vector whose
  entries are the eigenvalues of the Hermitian matrix $A$, arranged
  into non-increasing order.
\end{theorem}
Combining the results of Ky Fan and Thompson, we have the
following:
\begin{cor} \label{cor:Thompson-Fan}
  Let $H,K$ be Hermitian matrices.  Then there exists a Hermitian
  matrix $L$ such that
\be
  e^{i H} e^{i K} = e^{iL}; \,\,\,\,
  \lambda(L) \prec \lambda(H)+\lambda(K).
\ee
\end{cor}
\noindent
We will not apply this corollary directly, but we have included it
here because it captures the spirit of our later argument, combining
the Thompson and Ky Fan theorems to relate the properties of a product
of unitaries to the individual unitaries themselves.
Corollary~\ref{cor:Thompson-Fan} can be regarded as a vector-valued
analogue of the chaining property for dynamic strength measures used
in \cite{Nielsen03a} to establish lower bounds on computational
complexity.

\subsection{The canonical decomposition of a two-qubit gate}
\label{subsec:canonical}

%
%
The \emph{canonical decomposition} is a useful representation theorem
characterizing the non-local properties of a two-qubit unitary operator.
It was proved by Khaneja, Brockett, and Glaser~\cite{Khaneja01a} using
ideas from Lie theory.  Kraus and Cirac~\cite{Kraus01a} have given a
constructive proof using elementary notions, while Zhang \emph{et
al.}~\cite{Zhang02b} have discussed the decomposition in detail from the
point of view of Lie theory.  The decomposition states that any two-qubit
unitary $U$ may be written in the form
\begin{equation} \label{eq:can-decomp}
  U=(A_1\tp B_1)e^{i(\theta_x X\tp X+\theta_y Y\tp Y+\theta_z Z\tp Z)}
    (A_2\tp B_2),
\end{equation}
where $A_1$, $A_2$, $B_1$, $B_2$ are single-qubit unitaries, and the
three parameters, $\theta_x$, $\theta_y$, and $\theta_z$ characterize
the non-local properties of $U$.\footnote{Prior to~\cite{Khaneja01a},
  Makhlin~\cite{Makhlin00a} gave a proof that the non-local properties of
  $U$ are completely characterized by $\theta_x$, $\theta_y$ and
  $\theta_z$, but did not write down the canonical decomposition
  explicitly.}
Without loss of generality, we may choose the local unitaries to ensure
that
\begin{equation}
  \frac{\pi}{4} \ge \theta_x \ge \theta_y \ge |\theta_z|, \label{eq:order}
\end{equation}
and we refer to the set of parameters chosen in this way as the
\emph{canonical parameters} for $U$.  We will see below that these
parameters are unique. We define the {\em canonical form} of $U$ to be
\be
  U_c := (A_1^\dag\tp B_1^\dag)U(A_2^\dag\tp B_2^\dag);
\ee
up to local unitaries, $U_c$ is equivalent to $U$.  It will be
convenient to assume through the remainder of this section that $U$
has unit determinant.  This is equivalent to requiring that $A_1, A_2,
B_1, B_2$ can all be chosen to have unit determinant.

%
%
The canonical parameters turn out to be crucial to results about
simulation of two-qubit gates.  If
\be
  U_c = e^{i(\theta_x X\tp X + \theta_y Y\tp Y + \theta_z Z\tp Z)}
\ee
is the canonical form of $U$, then we define the \emph{non-local
content}, $\phi(U)$, of $U$ by $\phi(U) := \lambda(H_U)$, where
\be
  H_U := \theta_x \, X\tp X + \theta_y \, Y\tp Y + \theta_z \, Z\tp Z.
\ee
Explicitly, the components of $\phi(U)$ are
\begin{eqnarray}
\phi_1 &=&  \theta_x+\theta_y-\theta_z \label{eq:phi1} \\
\phi_2 &=&  \theta_x-\theta_y+\theta_z \label{eq:phi2} \\
\phi_3 &=& -\theta_x+\theta_y+\theta_z \label{eq:phi3} \\
\phi_4 &=& -\theta_x-\theta_y-\theta_z. \label{eq:phi4}
\end{eqnarray}

%
%
We now outline a simple procedure to determine the canonical
parameters of a two-qubit unitary operator.  Our explanation initially
follows~\cite{Hammerer02a} and~\cite{Leifer02a}.  However, as
explained below, there is an ambiguity in the procedure described in
those papers, related to the fact that the logarithm function has many
branches.  Our procedure resolves this ambiguity.

%
%
To explain the procedure, we need to introduce a piece of
notation, and explain a simple observation about single-qubit unitary
matrices.  The \emph{spin flip} operation on an arbitrary two-qubit
operator is defined as
\be
  \tilde M := (Y \tp Y) M^T (Y \tp Y),
\ee
where $Y$ is the Pauli sigma $y$ matrix, and the transpose operation
is taken with respect to the computational basis.  Note that the spin
flip operation may also be written as $\tilde M = M^T$, where the
transpose is taken with respect to a different basis, the \emph{magic
basis}~\cite{Hill97a},
\begin{eqnarray}
 \frac{|00\>+|11\>}{\sqrt{2}}; & &
i\frac{|00\>-|11\>}{\sqrt{2}}; \nonumber \\
i\frac{|01\>+|10\>}{\sqrt{2}}; & &
 \frac{|01\>-|10\>}{\sqrt{2}}.
\end{eqnarray}
The observation about single-qubit unitary matrices that we need is
the following.  Let $U$ be any single-qubit unitary matrix with unit
determinant.  Then
\begin{eqnarray} \label{eq:single-qubit-identity}
  U Y U^T = Y,
\end{eqnarray}
where the transpose is taken in the computational basis.  This simple
identity is easily verified.

%
%
Now suppose $U$ is an arbitrary two-qubit unitary with unit determinant.
By definition of the spin flip, and substituting the canonical
decomposition, we have
\begin{eqnarray}
  U \tilde U & = & (A_1\tp B_1) U_c (A_2\tp B_2) (Y \tp Y)  \nn
         && \times (A_2^T \tp B_2^T) U_c (A_1^T \tp B_1^T) (Y \tp Y).
\end{eqnarray}
By the identity Eq.~(\ref{eq:single-qubit-identity}) we see that
\begin{eqnarray}
  U \tilde U = (A_1\tp B_1) U_c (Y \tp Y) U_c (A_1^T \tp B_1^T) (Y \tp Y).
\end{eqnarray}
Using the fact that $Y \tp Y$ commutes with $X \tp X$, $Y \tp Y$, and $Z
\tp Z$, we see that $Y \tp Y$ commutes with $U_c$, and thus
\begin{eqnarray}
  U \tilde U = (A_1\tp B_1) U_c^2 (Y \tp Y)(A_1^T \tp B_1^T) (Y \tp Y).
\end{eqnarray}
Finally, applying Eq.~(\ref{eq:single-qubit-identity}) again gives
\begin{eqnarray} \label{eq:almost-can-decomp}
  U \tilde U = (A_1\tp B_1) U_c^2 (A_1^\dagger \tp B_1^\dagger)
\end{eqnarray}
Eq.~(\ref{eq:almost-can-decomp}) suggests a procedure to determine the
canonical parameters for $U$, based on the observation that \be
\label{eq:can-eig-decomp} \lambda(U \tilde U) = \lambda(U_c^2) =
(e^{2i \phi_1}, e^{2i \phi_2}, e^{2i\phi_3},e^{2i\phi_4}), \ee where
the $\phi_j$ are related to the canonical parameters
$\theta_x,\theta_y$ and $\theta_z$ by
Eqs.~(\ref{eq:phi1})--(\ref{eq:phi4}).  It is tempting to conclude
that one can determine $\theta_x,\theta_y,\theta_z$ from the
eigenvalues of $U \tilde U$, simply by taking logarithms and inverting
the resulting linear equations.  Indeed, such a conclusion is reached
in~\cite{Hammerer02a} and~\cite{Leifer02a}, using arguments similar to
those just described.  Unfortunately, determining the canonical
parameters is not quite as simple as this, because $z \rightarrow
e^{iz}$ is not a uniquely invertible function.  In particular, $e^{iz}
= e^{i(z+2\pi m)}$, where $m$ is any integer, so there is some
ambiguity about which branch of the logarithm function to use in
calculating the canonical parameters.  In fact, we prove later that no
one branch of the logarithm function can be used.  However, these
considerations do allow us to reach the following conclusion:
\begin{lemma}
\label{lemma:can-formula}
  Let $U$ be a two-qubit unitary.  Then there exists a Hermitian $H$
  such that
  \begin{eqnarray}
    U \tilde U = e^{2iH}, \,\,\,\, \lambda(H) = \phi(U).
  \end{eqnarray}
  Moreover, if $H$ is any Hermitian matrix such that $\lambda(U \tilde
  U) = \lambda(e^{2iH})$ then it follows that $\lambda(H) = \phi(U) +
  \pi \vec m$, where $\vec m$ is some vector of integers.
\end{lemma}
Although this lemma is sufficient to prove our later results, there is
in fact a simple method for exactly calculating the canonical
parameters.  Because there are many applications of the canonical
decomposition, we describe this method in the appendix.  The method
will not be needed elsewhere in the paper.

\subsection{The canonical form of a two-qubit Hamiltonian}
\label{subsec:canonicalham}

Finally, we introduce one additional concept, the \emph{canonical
  form} of a two-qubit Hamiltonian, $H$ \cite{Dur01a}.  Any two-qubit
Hamiltonian $H$ can be expanded as
\be
  H = \sum_{j,k=0}^3 h_{jk} \, \sigma_j \tp \sigma_k.
\ee
Then let
\be
  H' := {H+\tilde H \over 2}
      = \sum_{j,k \neq 0} h_{jk} \, \sigma_j \tp \sigma_k.
\ee
That is, $H'$ is just the Hamiltonian that results when the local
terms in $H$ are removed.  It is not difficult to show that $H$ and
$H'$ are interchangeable resources for simulation in the sense that,
given fast local unitaries, evolution according to $H$ for a time $t$
can be simulated by evolution according to $H_c$ for a time $t$, and
vice versa.  Furthermore, by doing appropriate local unitaries, it can
be shown~\cite{Dur01a} that simulating $H'$ (and thus $H$) is
equivalent to simulating the canonical form of $H$,
\begin{eqnarray}
  H_c = h_x \, X\tp X + h_y \, Y\tp Y + h_z \, Z\tp Z,
\end{eqnarray}
where $h_x \geq h_y \geq |h_z|$.  Once again, $H$ and $H_c$ are
interchangeable resources for simulation.

Note that the three parameters $h_x,h_y,h_z$ are completely
characterized by the three degrees of freedom in
$\lambda(H_c)=\lambda(H+\tilde H)/2$, just as the three canonical
parameters $\theta_x,\theta_y,\theta_z$ are completely characterized
by the three degrees of freedom in $\lambda(U_c^2)=\lambda(U \tilde
U)$.

\section{Simulation of two-qubit gates}
\label{sec:two-qubit}

%
%
We now return to the main purpose of the paper, proving results about
the time to simulate a unitary gate using entangling Hamiltonians and
fast local gates.  We aim to prove the following result:
\begin{theorem}[Vidal, Hammerer, Cirac \cite{Vidal02a,Hammerer02a},
cf. Khaneja, Brockett and Glaser~\cite{Khaneja01a}]
\label{thm:VHC}
  Let $U$ be a two-qubit unitary operator, and let $H$ be a two-qubit
  entangling Hamiltonian.  Then the minimal time required to simulate
  $U$ using $H$ and fast local unitaries is the minimal value of $t$
  such that there exists a vector of integers $\vec m$ satisfying
  \begin{eqnarray} \label{eq:2-constraint}
    \phi(U) + \pi \vec m \prec \frac{\lambda(H+\tilde H)}{2} \, t.
  \end{eqnarray}
\end{theorem}
\noindent
Note further that only two vectors of integers need to be checked,
$\vec m = (0,0,0,0)$ and $\vec m = (1,1,-1,-1)$, since all the other
possibilities give rise to weaker constraints on the minimal time, $t$
\cite{Vidal02a,Hammerer02a}.  The difficult part of the proof of
Theorem~\ref{thm:VHC} is the proof that Eq.~(\ref{eq:2-constraint}) is
a lower bound on the simulation time, $t$, and it is this part of the
proof that we focus on simplifying.  The proof that this lower bound
may be achieved follows from standard results on majorization, and we
refer the interested reader to~\cite{Vidal02a,Hammerer02a} for
details.

%
%
To prove that Eq.~(\ref{eq:2-constraint}) constrains the minimal time
for simulation, we begin by characterizing the canonical decomposition
of a product of unitary matrices.  Let $\Lambda(U) := \lambda(U\tilde
U)$, and define the equivalence relation $A \sim B$ for Hermitian
matrices $A$ and $B$ iff $\lambda(A) = \lambda(B)$.  Then we have:
\begin{lemma}
\label{lemma:2-proof}
  Let $U_j$ be unitary matrices, and let $H_j$ be Hermitian matrices such
  that $U_j \tilde U_j = e^{2iH_j}$.  Then there exist Hermitian
  matrices $K_j$ such that $H_j \sim K_j$, and
\be
  \Lambda(U_N \ldots U_1) = \lambda(e^{2i(K_1+\cdots+K_N)}).
\label{eq:2-proof}
\ee
\end{lemma}

\begin{proof}
We induct on $N$.  The result is trivial for $N = 1$, so we need only
consider the inductive step.
Using the fact $\lambda(AB) = \lambda(BA)$, we have
\be
  \Lambda(U_{N+1}\ldots U_1) =
  \lambda(\tilde U_{N+1} U_{N+1} \, U_N \ldots U_1 \tilde U_1 \ldots
  \tilde U_N).
\ee
By the inductive hypothesis there exist Hermitian $K_j'$ such that
$H_j \sim K_j'$ and
\be
  \lambda(U_N \ldots U_1 \tilde U_1 \ldots \tilde U_N) =
  \lambda(e^{2i(K_1'+\cdots+K_N')}).
\ee
Therefore, $U_N \ldots U_1 \tilde U_1 \ldots \tilde U_N
  = e^{2i(K_1''+\cdots+K_N'')}$,
for some $K_j'' \sim H_j$.  Observe also that
\be
  \tilde U_{N+1}U_{N+1} \sim U_{N+1}\tilde U_{N+1}
                           = e^{2iH_{N+1}},
\ee
and thus
$\tilde U_{N+1}U_{N+1} = e^{2iK_{N+1}''}$
for some $K_{N+1}'' \sim H_{N+1}$.  It follows by substitution that
\be
  \Lambda(U_{N+1}\ldots U_1) =
  \lambda(e^{2i K_{N+1}''} e^{2i(K_1''+\cdots+K_N'')}).
\ee
Applying Thompson's theorem gives
\be
  \Lambda(U_{N+1}\ldots U_1) = \lambda(e^{2i (K_1+\cdots+K_{N+1})})
\ee
for some $K_j \sim K_j'' \sim H_j$,
which completes the inductive step of the proof.
\end{proof}

Given this result, it is straightforward to complete the proof of
Eq.~(\ref{eq:2-constraint}).

\begin{proof}
Write $U$ in the form
\begin{eqnarray}
  U = e^{-i H t_1} V_1 e^{-i H t_2} V_2 \ldots V_{k-1} e^{-iH t_k},
\end{eqnarray}
where $t_1,\ldots,t_k$ are times of evolution, $t = t_1 + \ldots +
t_k$ is the total time for simulation, and $V_j$ are local unitaries.
Without loss of generality, we may assume $H$ is in canonical form.
Applying Lemma~\ref{lemma:2-proof}, we obtain
\begin{eqnarray}
  \Lambda(U) = \lambda(e^{2i(H_1 t_1+\ldots + H_k t_k)})
\label{eq:Lambdau}
\end{eqnarray}
where $H_j \sim H$ for each $j$.  Here we have used the observation $V_j
\tilde V_j = 1$, so all the contributions from local unitaries vanish.  It
follows from Lemma~\ref{lemma:can-formula} that
\begin{eqnarray}
  \phi(U)+\pi \vec m = \lambda(H_1 t_1+\ldots + H_m t_m),
\end{eqnarray}
and using Ky Fan's theorem gives
\begin{eqnarray}
  \phi(U)+\pi \vec m \prec \lambda(H)(t_1+\ldots +t_m),
\end{eqnarray}
which is Eq.~(\ref{eq:2-constraint}), as desired.
\end{proof}

\section{Discussion}
\label{sec:conc}

In this paper, we have provided a simplified proof of a lower bound on
the time required to simulate a two-qubit unitary gate using a given
two-qubit interaction Hamiltonian and local unitaries.  The bound
follows easily from standard results on majorization together with
Thompson's theorem on products of unitary operators.

Although we have described canonical decompositions of two-qubit gates
in some detail, we note that our proof does not actually require
properties of the decomposition unique to two qubits.  In fact, it is
straightforward to prove an analogue of Eq.~(\ref{eq:2-constraint})
for an $n$-qubit system.  For an $n$-qubit operator $M$, suppose we
define a generalized spin flip $M \rightarrow \tilde M$, where $\tilde
M$ is the transpose operation in a basis such that, whenever $M$ is
local, $M$ is orthogonal, i.e., $M\tilde M = I$.  It is not difficult
to construct examples of such bases, at least when $n$ is even.  An
example is the basis obtained by rotating the computational basis
using the transformation $(I-iY^{\otimes n})/\sqrt 2$, for $n$ even.
This basis change gives $\tilde M = Y^{\otimes n} M^T Y^{\otimes n}$,
where the transpose is taken in the computational basis, and thus this
operation generalizes the transpose in the magic basis.  In this
general setting the following lower bound on the time required to
implement an $n$-qubit gate holds:
\begin{cor}
  Let $U$ be an $n$-qubit unitary operator, and let $H$ be an
  $n$-qubit Hamiltonian.  Then the time required to simulate $U$ using
  $H$ and fast local unitaries satisfies
  \begin{eqnarray}
    {1 \over 2} \arg \lambda(U \tilde U) + \pi \vec m
    \prec \frac{\lambda(H+\tilde H)}{2} \, t.
  \label{eq:n-constraint}
  \end{eqnarray}
  for some vector of integers $\vec m$.
\end{cor}
\noindent
The proof follows simply by taking the arguments of both sides of
Eq.~(\ref{eq:Lambdau}) and applying Ky Fan's theorem.  All steps leading
up to Eq.~(\ref{eq:Lambdau}) remain valid for $n$-qubit systems using the
above definition of the generalized spin flip.

Unfortunately, we have not found any interesting examples with $n>2$
for which Eq.~(\ref{eq:n-constraint}) provides a nontrivial lower
bound on the time required to implement some quantum gate.  It would
be interesting to construct cases where Eq.~(\ref{eq:n-constraint})
(or some similar condition) does give a nontrivial constraint on
multipartite gate simulation.  One might imagine that such techniques
could be used to prove circuit lower bounds on certain quantum
computations, although it does not seem likely that such bounds would
be especially strong, given the well-known difficulty of this problem.

\acknowledgments

We thank Aram Harrow and Tobias Osborne for helpful discussions, and
Andrew Doherty for an informative seminar on related problems.
AMC received support from the Fannie and John Hertz Foundation, and
thanks the University of Queensland node of the Centre for Quantum
Computer Technology for its hospitality.  AMC was also supported in
part by the Cambridge--MIT Institute, by the Department of Energy
under cooperative research agreement DE-FC02-94ER40818, and by the
National Security Agency and Advanced Research and Development
Activity under Army Research Office contract DAAD19-01-1-0656.
Finally, we acknowledge the hospitality of the Caltech Institute for
Quantum Information, where this work was completed.  This work was
supported in part by the National Science Foundation under grant
EIA-0086038.

\appendix
\section*{Appendix: A method for computing the canonical parameters
          of a two-qubit unitary gate}

In this appendix, we describe a method for computing the canonical
parameters of a two-qubit unitary, based on the discussion in
Section~\ref{subsec:canonical}.  The key is to take logarithms in just
the right way.  From Eqs.~(\ref{eq:order})
and~(\ref{eq:phi1})--(\ref{eq:phi4}), we see that
\begin{equation}
  \frac{3\pi}{2} \ge 2\phi_1 \ge 2\phi_2 \ge 2\phi_3 \ge 2\phi_4
                 \ge -\frac{3\pi}{2}.
\label{eq:phiorder}
\end{equation}
It is not difficult to find examples where the first or last
inequality is saturated, so no single fixed branch of the logarithm
function can be used to determine the $\phi_j$.  One might hope
instead that there exists a method for choosing a different branch for
each particular $U$, so that the corresponding $2\phi_j$ lie within
that branch.  However, even this is not possible in general. To
understand this, note that
\begin{equation}
2\phi_1-2\phi_4=4(\theta_x+\theta_y).  \label{eq:thetaxthetay}
\end{equation}
In cases where $\theta_x=\theta_y=\pi/4$, we have $2\phi_1-2\phi_4=2\pi$,
in which case the values $2\phi_j$ do not lie in {\em any} one branch.

We now show how to compute the $\phi_j$.  The idea is that we can first
take the argument of the eigenvalues in Eq.~(\ref{eq:can-eig-decomp}) over
some fixed branch.  Then we can systematically determine which of the
resulting values have been shifted by $2\pi$ from the value $2\phi_j$ (due
to an incorrect branch) and correct these values accordingly.

Let $S_j$, $j=1,\dots,4$ be defined as follows:
\begin{equation}
  2S_j=\arg (e^{2i\phi_j}).
\end{equation}
That is, $2S_j$ are the arguments of the eigenvalues of $U\tilde U$,
where we take the argument over the branch $(-\frac{\pi}{2},
\frac{3\pi}{2}]$, so that the $S_j$ are contained in the interval
$(-\frac{\pi}{4}, \frac{3\pi}{4}]$.  Considering the range of values
that $\phi_j$ may take, from Eq.~(\ref{eq:phiorder}), and the
particular branch we are using, it is clear that:
\begin{equation}
  S_j= \left\{
  \begin{array}{lcl}
  \phi_j+\pi &\hspace{0.5cm}&
  \mbox{if } \phi_j\le-\frac{\pi}{4} \\
  \phi_j & & \mbox{otherwise.}
  \end{array}
  \right. \label{eq:sj}
\end{equation}
{}From Eqs.~(\ref{eq:phi1})--(\ref{eq:phi4}) we have
\begin{equation}
  \phi_1+\phi_2+\phi_3+\phi_4=0. \label{eq:sumphi}
\end{equation}
Combining Eqs.~(\ref{eq:sj}) and~(\ref{eq:sumphi}), we see
that
\begin{equation}
  S_1+S_2+S_3+S_4=\pi n,
\end{equation}
where $n$ is the number of $\phi_j$ that are less than or equal to
$-\frac{\pi}{4}$. Possible values for $n$ are $0, 1, 2$ and $3$ (all
four $\phi_j$ cannot simultaneously be $\le-\frac{\pi}{4}$, since that
would contradict Eq.~(\ref{eq:sumphi})). Since the $\phi_j$ obey the
ordering in Eq.~(\ref{eq:phiorder}), then the $n$ values of $\phi_j$
that are less than or equal to $-\frac{\pi}{4}$ are
$\phi_4,\dots,\phi_{4-n+1}$, and the remaining $4-n$ values greater
than $\frac{\pi}{4}$ are $\phi_1,\dots,\phi_{4-n}$.  Thus, using
Eq.~(\ref{eq:sj}), we see that the set of values $S_j$ consist of $n$
``shifted'' $\phi_j$ values
\begin{equation}
  \phi_4+\pi,\dots,\phi_{4-n+1}+\pi, \label{eq:shifted}
\end{equation}
and $4-n$ ``non-shifted'' values of $\phi_j$
\begin{equation}
  \phi_1,\dots,\phi_{4-n}.  \label{eq:nonshifted}
\end{equation}
Furthermore, all of the shifted values in (\ref{eq:shifted}) are no
less than any of the non-shifted values in (\ref{eq:nonshifted}).
This is shown by combining Eq.~(\ref{eq:order}) with
Eq.~(\ref{eq:thetaxthetay}), giving $\phi_1-\phi_4 \le \pi$, which
when combined with Eq.~(\ref{eq:phiorder}) implies that $\phi_j \le
\phi_k + \pi$ for all $j,k$, as required. Therefore, the largest $n$
values of $S_j$ are guaranteed to be the values in (\ref{eq:shifted}).
Thus subtracting $\pi$ from the largest $n$ values of $S_j$, gives us
$\phi_4,\dots,\phi_{4-n+1}$, and the the remaining $4-n$ values of
$S_j$ give us $\phi_1,\dots,\phi_{4-n}$.

In summary, the nonlocal parameters $\theta_x, \theta_y$ and
$\theta_z$ may be computed as follows. Find the arguments of the
eigenvalues of $U\tilde U$ over the branch $(-\frac{\pi}{2},
\frac{3\pi}{2}]$. Call these values $2S_j$. Calculate
$n=(S_1+S_2+S_3+S_4)/\pi$.  Replace the $n$ largest values of $S_j$ by
those values minus $\pi$.  The resulting values, when placed in
nonincreasing order, are equal to $(\phi_1,\phi_2,\phi_3,\phi_4)$. The
parameters $\theta_x,\theta_y$ and $\theta_z$ are then found by
inverting Eqs.~(\ref{eq:phi1})--(\ref{eq:phi4}).

\bibliography{mybib}

\end{document}